\documentclass[reprint,superscriptaddress,amsmath,amssymb, aps,prl,floatfix,]{revtex4-1}

\pdfoutput=1 
\usepackage{graphicx}
\usepackage{bm}
\usepackage{hyperref}
\usepackage{times}

\begin{document}

\title{NMR Study of the Superconducting Gap Variation near the Mott Transition in Cs$_{3}$C$_{60}$}

\author{P. Wzietek}
\affiliation{Laboratoire de Physique des Solides, Universit\'e Paris-Sud 11, CNRS UMR 8502, 91405 Orsay, France}
 \email{pawel.wzietek@u-psud.fr}
\author{T. Mito}
\affiliation{Laboratoire de Physique des Solides, Universit\'e Paris-Sud 11, CNRS UMR 8502, 91405 Orsay, France}
\affiliation{Graduate School of Material Science, University of Hyogo, Kamigori, Hyogo 678-1297, Japan}

\author{H. Alloul}
\affiliation{Laboratoire de Physique des Solides, Universit\'e Paris-Sud 11, CNRS UMR 8502, 91405 Orsay, France}

\author{D.~Pontiroli}
\affiliation{Dipartimento di Fisica e Scienze della Terra, Universit{\`a} di Parma - Via G.P.Usberti 7/a, I-43124 Parma, Italy}
\affiliation{Laboratorio MIST.E-R, P. Gobetti 101, I-40129 Bologna, Italy}
\author{M. Aramini}
\author{ M.~Ricc{\`o}}
\affiliation{Dipartimento di Fisica e Scienze della Terra, Universit{\`a} di Parma - Via G.P.Usberti 7/a, I-43124 Parma, Italy}

\begin{abstract}
Former extensive studies of superconductivity in the \textit{A}$_{3}$C$_{60}$ compounds,
where \textit{A} is an alkali metal, have led one to consider that Bardeen-Cooper-Schrieffer (BCS) electron-phonon pairing prevails in those compounds, though
the incidence of electronic Coulomb repulsion has been highly debated. The
discovery of two isomeric fulleride compounds Cs$_{3}$C$_{60}$ which exhibit
a transition with pressure from a Mott insulator (MI) to a
superconducting (SC) state clearly reopens that question. Using pressure ($p$) as a
single control parameter of the C$_{60}$ balls lattice spacing, one can now
study the progressive evolution of the SC properties when the electronic
correlations are increased towards the critical pressure $p_{c}$ of the Mott
transition. We have used $^{13}$C and $^{133}$Cs NMR measurements on the
cubic phase A15-Cs$_{3}$C$_{60}$ just above $p_{c}=5.0(3)$ kbar, where the
SC transition temperature $T_{c}$ displays a dome shape with decreasing cell
volume.  From the $T$ dependence below $T_{c}$ of the nuclear spin lattice
relaxation rate $(T_{1})^{-1}$ we determine the electronic excitations in the
SC state, that is $2\Delta$, the gap value.
 The latter is found to be largely enhanced with respect to the BCS value established in the case
  of dense \textit{A}$_{3}$C$_{60}$ compounds. 
It even increases slightly with decreasing $p$ towards $p_{c}$, where $T_{c}$ decreases on
the SC dome, so that $2\Delta /k_{B}T_{c}$ increases regularly upon
approaching the Mott transition. These results bring clear evidence that
the increasing correlations near the Mott transition are not significantly
detrimental to superconductivity. They rather suggest that repulsive electron interactions
might even reinforce elecron-phonon superconductivity, being then partly responsible for
the large $T_{c}$ values, as proposed by theoretical models taking the
electronic correlations as a key ingredient.

\begin{description}
\item[PACS numbers] 
74.70.Wz, 
71.30.+h,   
74.25nj   
\end{description}
\end{abstract}

\maketitle

Since the BCS proposal  it has always been clear that, in order to drive a
superconducting (SC) state, the attractive interaction between electrons
mediated by phonons must overcome the Coulomb repulsion between the
electrons. Conversely, if the repulsion is strong
enough, as in the cuprate high temperature superconductors (HTSC), it tends to
localize the electrons on atomic sites, leading to a Mott insulator
(MI) that may be magnetically ordered at low temperature. In the
cuprates the appearance of superconductivity is  investigated by doping such a magnetic MI. 
Many other systems (pnictides, heavy fermions) display phase diagrams in which a metallic magnetic state is proximate to a SC state.
It is now recognized that AF fluctuations
might sometimes mediate the SC pairing in such systems, rather than 
 electron-phonon coupling. By contrast, the absence of nearby
magnetic phases in a SC family is often considered a sign of conventional
superconductivity with negligible incidence of Coulomb repulsion. 

This scenario has been initally applied to the HTSC alkali-fulleride salts 
\textit{A}$_{3}$C$_{60}$ discovered only a few years after the cuprates 
\cite{gunnarsson-RMP69}. However, the fact that $A_{n}$C$_{60}$ compounds with
even $n=2,4$  are nonmagnetic insulators has always suggested that Coulomb
repulsions are important in the fulleride families. Furthermore the strength
of electronic correlations is in this case reinforced  by Jahn-Teller distortions
(JTD) of the C$_{60}$ ball, which localize the electrons on the balls in
singlet nonmagnetic (low-spin) states \cite{Brouet-Springer2004,Fabrizio-PRB55}. 
A significant effort to determine whether \textit{A}$_{3}$C$_{60}$ compounds 
could become MI by increasing the separation between C$_{60}$
balls has been successful recently, thanks to the synthesis of the
compound with \textit{A}=Cs, the alkali metal with the largest ionic radius
 \cite{Ganin-NMAT7,Takabayashi-Science323,Ihara-PRL104}. This expanded fulleride
 Cs$_{3}$C$_{60}$ being highly compressible, it has been possible to recover a
SC state by application of pressure, without charge modification,
by analogy with the situation in layered organic compounds \cite{Lefebvre00prl,Kagawa05nmat}. 
The phase diagram looks then very much like those of  the other families of
unconventional correlated electron superconductors. This opens an original possibility to
study the evolution from a MI to a SC state in this three-dimensional
compound. 

As two isomeric phases of Cs$_{3}$C$_{60}$ could be synthesized, the initial
studies have been mostly focused on the differences in their respective 
($T,p $) phase diagrams and of their magnetic properties in the Mott state, in
which similar dynamic JTD have been detected \cite{Klupp2012nat}. But so far
very little has been achieved concerning the incidence of the growth of the
electronic correlations on both the microscopic SC and metallic properties
when approaching the Mott transition.

We address this question here by the use of $^{13}$C and $^{133}$Cs NMR,
namely the spin lattice $(T_{1})^{-1}$ measurements which uniquely permit
us to determine the SC gap magnitude \cite{pennington-RMP68}
through the pressure $p_{c}$ of the MI to SC transition. We first confirm that
the $s$-wave symmetry of the order parameter is maintained, down to $p_{c}$.
We find evidence, however, that near $p_{c}$ the SC gap does not follow the
BCS weak coupling scenario which appeared to prevail in dense 
fcc-$A_{3}$C$_{60}$ (e.g. \textit{A}=K, Rb) \cite{pennington-RMP68,Ricco-PhC98,Kiefl-PRL70}.
 Also, at variance with the
pseudogap behaviour found in underdoped cuprates \cite{Alloul89prl}, the
spin susceptibility exhibits a regular increase with correlations, up to the
Mott transition. All these features might be related to a quite original SC
pairing mechanism \cite{Capone-Science296}, local in nature since mediated by on-ball optical phonon
modes. Indeed, a new paradigm for superconductivity in fullerides has been
conceived using dynamical mean field theory (DMFT) calculations which show
that such a pairing scheme could in fact be reinforced by electronic
correlations: a SC state with an enhanced $T_{c}$ in the vicinity of the
Mott transition has been anticipated \cite{Capone-Science296,Capone-RMP81}. 

\paragraph{Experimental techniques and sample characterization.}

\begin{figure}[tb]
\begin{center}
\includegraphics[width=0.99\hsize]{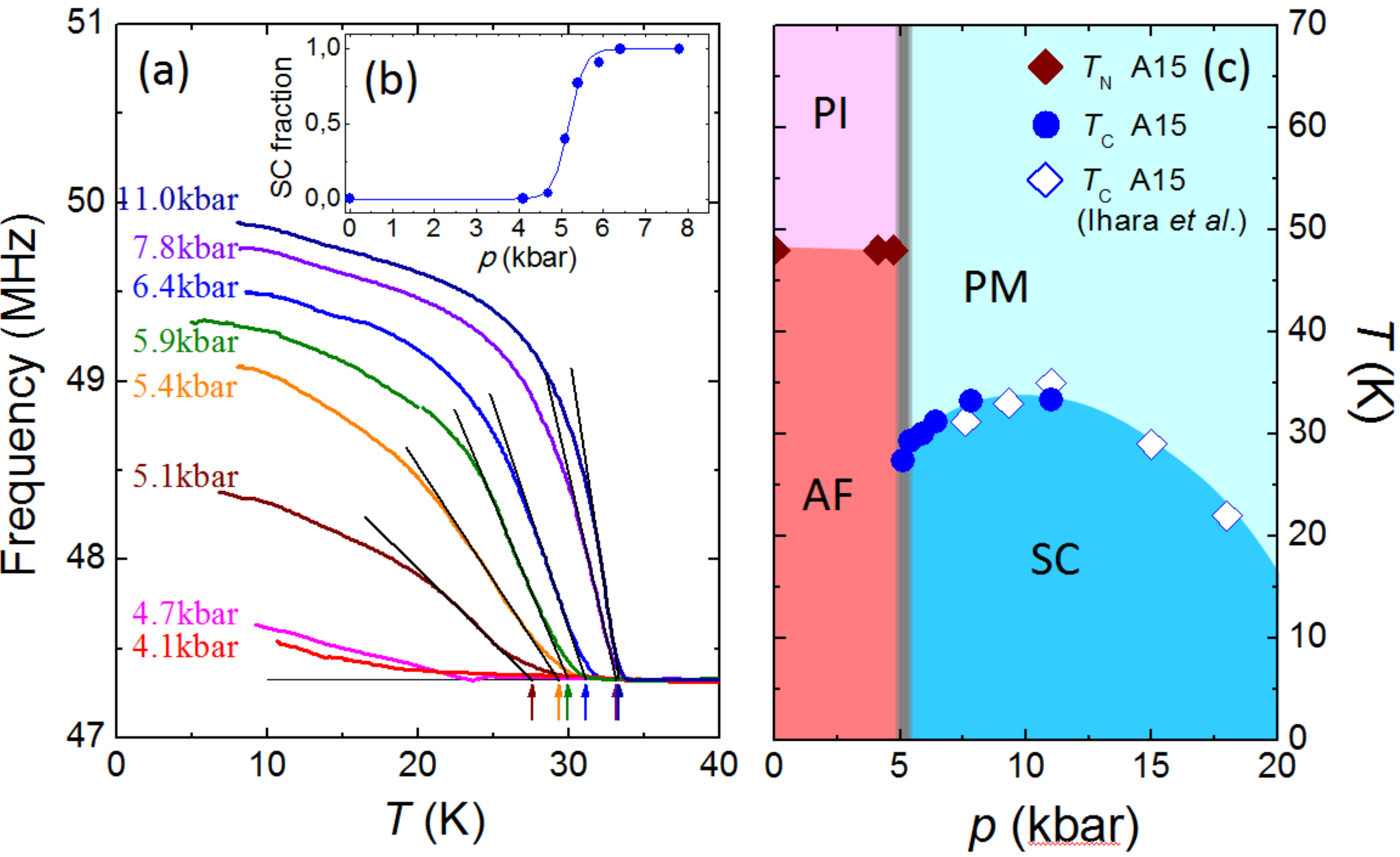}
\end{center}
\caption{(a) The zero field $T$ dependence of the resonance
frequency of the NMR\ coil monitors the growth of the SC diamagnetism with
increasing  $p$. (b) The $p$ variation of the low $T$  diamagnetic signal
permits us to locate the Mott transition at $p_{c}=5\pm 0.3$ kbar. 
(c) The $T_{c}(p)$ phase diagram determined for this sample is completed by  higher $p$ data taken on a former A15 phase sample \cite{Ihara-PRL104}.
}
\label{Diamagnetism}
\end{figure}

Improved synthesis techniques \cite{Ganin2010polymorphism} allowed us to
produce a sample with 
 an 85\%  ratio of A15/fcc phases (A15 73\%, fcc 15\%, bco 12\%), much larger than that used previously  \cite{Ihara-PRL104}. 
A large amount of this batch has been
introduced in the NMR coil in a glove box, and sealed there in a high
pressure cell designed for NMR experiments. The zero field variation of the
coil tuning frequency monitors the SC diamagnetic response of the sample.
The data of Fig.\ref{Diamagnetism}(a) have been collected at increasing $p$,
in the same set up. The onset of superconductivity occurs quite sharply with increasing 
$p$, as monitored in Fig.\ref{Diamagnetism}(b). A full SC state is achieved above 5.9
kbar, while the residual signal at 4.1 kbar is assigned to the small fcc
phase content which has a lower $p_{c}$value. \cite{Ihara-PRL104}. The large
A15 fraction  in this sample allowed a more precise determination of the
 $T_{c}(p)$ phase diagram [Fig.\ref{Diamagnetism}(c)] than in former reports 
\cite{Takabayashi-Science323,Ihara-PRL104}.

\paragraph{NMR spectra and SC state spin susceptibility near $p_{c}$.}

\begin{figure}[tb]
\begin{center}
\includegraphics[width=0.99\hsize]{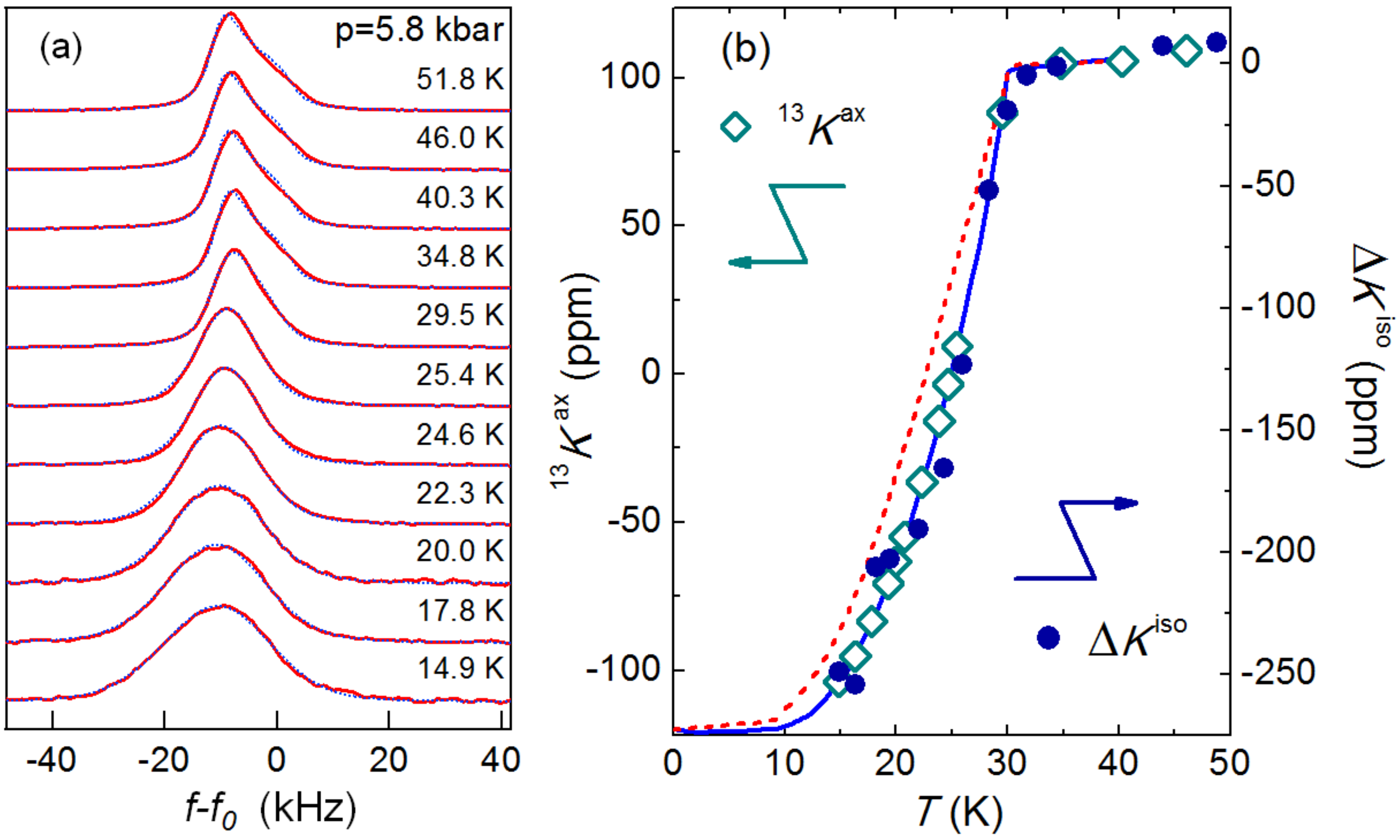}
\end{center}
\caption{ The $^{13}$C NMR spectra (measured at $f_0=82$ MHz)  at $p=$5.8 kbar (a)  exhibit
anisotropic line shapes, which evolve abruptly below $T_{c}$ with a change of
sign of the anisotropy $K^{ax}$.
 The fits of the spectra (dotted blue lines in (a), see \cite{sup})  are equivalently good if $K^{ax}(T)$ is governed by  Yosida function with $\Delta /k_{B}T_{c}=1.75$ (dotted red line in (b)) or with $\Delta /k_{B}T_{c}=2.2$ (solid blue line in (b)). In the latter case the imposed  $K^{ax}(T)$ (lozenges in (b)) is compared to the data obtained for $\Delta K^{iso}$  (see text and \cite{sup}).
}
\label{13C spectra}
\end{figure}

The $^{133}$Cs spectra of the A15 phase \cite{sup} were found identical to
those published formerly \cite{Jeglic-PRB80,Ihara-PRL104}. More important
for the study of the SC state properties are the $^{13}$C spectra displayed
in Fig.~\ref{13C spectra}(a), to be discussed below.

The NMR shift of a nuclear spin site is a tensor $K^{\alpha }$
 \cite{pennington-RMP68}:  
\begin{equation}
K^{\alpha }=K_{s}^{\alpha }(T)+K_{orb}^{\alpha }+K_{dia}(T).
\label{equ NMR shift}
\end{equation}
where $\alpha $ labels the orientation of the applied field $H$ relative to
the local crystallographic axis  and $K_{dia}$ is the inhomogeneous field reduction due to the
vortices and macroscopic screening currents in the SC state.  Here the Knight shift
 $K_{s}^{\alpha}=A^{\alpha }\chi _{s}(T)$ probes the electronic spin susceptibility
  $\chi_{s}(T)$, which vanishes at $T=0$ for a singlet SC state. 
  The usually $T$ independent $K_{orb}^{\alpha }$,
 due to orbital currents associated with
filled electronic shells, is the unique contribution to $K^{\alpha }$ in
nonmetallic $A_{n}$C$_{60}$ compounds, such as pure C$_{60}$
\cite{Tycko-PRL67},  Na$_{2}$C$_{60}$ or  K$_{4}$C$_{60}$\cite{Brouet-PRB66bis}. 

 In powder samples the shift of the first moment of the NMR spectrum is given by the isotropic
component $K^{iso}=(1/3)\sum K^{\alpha }$, while the spectral shape depends on the
anisotropic traceless contribution to $K^{\alpha }$. The latter dominates
for $^{13}$C \cite{pennington-RMP68,Ihara-PRL104}  and yields the
characteristic shape seen above $T_{c}=30$~K in Fig.\ref{13C spectra}(a). In
that case  $K^{\alpha }$ is nearly axial and a single component 
 $K^{ax}$ is required to characterize the spectral shape \cite{sup}. 

Spectra in the SC state  could only be taken for $p\geq5.8$ kbar for which the sample
displays bulk superconductivity below $T_{c}$, without any leftover from the Mott state.
For $^{13}$C one can see in Fig.\ref{13C spectra}(a) a fast variation of the spectrum 
shape below $T_{c}$ corresponding to a sign change of  $K^{ax}$.
This  is direct evidence that 
$\chi_{s}(T)$ drops down as expected for singlet $s$-wave superconductivity,
 as $K_{orb}^{ax}$ \cite{note-Korb} has an opposite sign to the normal state
value of $K^{ax}$. Quantitative estimates of $\chi _{s}(T)$ were possible thanks
 to the excellent fits of the $^{13}$C NMR spectra as shown in Fig.\ref{13C spectra}(a).
 They permit us to ensure the overall consistency
with the expected variation of $K_{s}^{ax}(T)$ for a BCS singlet state 
displayed in Fig.\ref{13C spectra}(b). The accuracy on  $K_{s}^{ax}(T)$ is, 
however, slightly hampered \cite{sup} as one needs to consider the spectrum
broadening induced by $K_{dia}$, which increases progressively below $T_{c}$,
 as seen in Fig.\ref{13C spectra}(a). 

An independent determination of $\chi _{s}(T)$ is available from the $^{13}$C 
and $^{133}$Cs  isotropic shifts $K^{iso}$, which display a significant
decrease below $T_{c}$, as reported in \cite{Ihara-PRL104} and in \cite{sup}.
 The fields induced by the screening currents being independent on the
nuclear probe, $K_{dia}$ is eliminated in $\Delta K^{iso}\equiv
^{133}K^{iso}-^{13}K^{iso}$, which therefore reflects the variation of
 $\chi_{s}(T)$ below $T_{c}$. 

In Fig.~\ref{13C spectra}(b) we show
  $\Delta K^{iso}$           
 \footnote{$\Delta K^{iso}$ has been determined from the 
 $^{13}$C data at $p=5.8 kbar$ and $^{133}$Cs data at $p=5.9 kbar$.
} 
 scaled to permit the best fit with $K^{ax}(T)$. We adapted the $\Delta K^{iso}$ scale to attempt fits with either the solid and dotted curves. We found that the sharp drop of $\Delta K^{iso}$  seen just below $T_c$ is better reproduced by scaling the experimental points with the  larger SC gap. Though this is not a fully secured conclusion this suggests that the drop of $\chi_{s}(T)$ might be sharper than the pure BCS function as in strong coupling superconductors.

\paragraph{SC gap from spin lattice relaxation data.}

\begin{figure}[tb]
\begin{center}
\includegraphics[width=0.99\hsize]{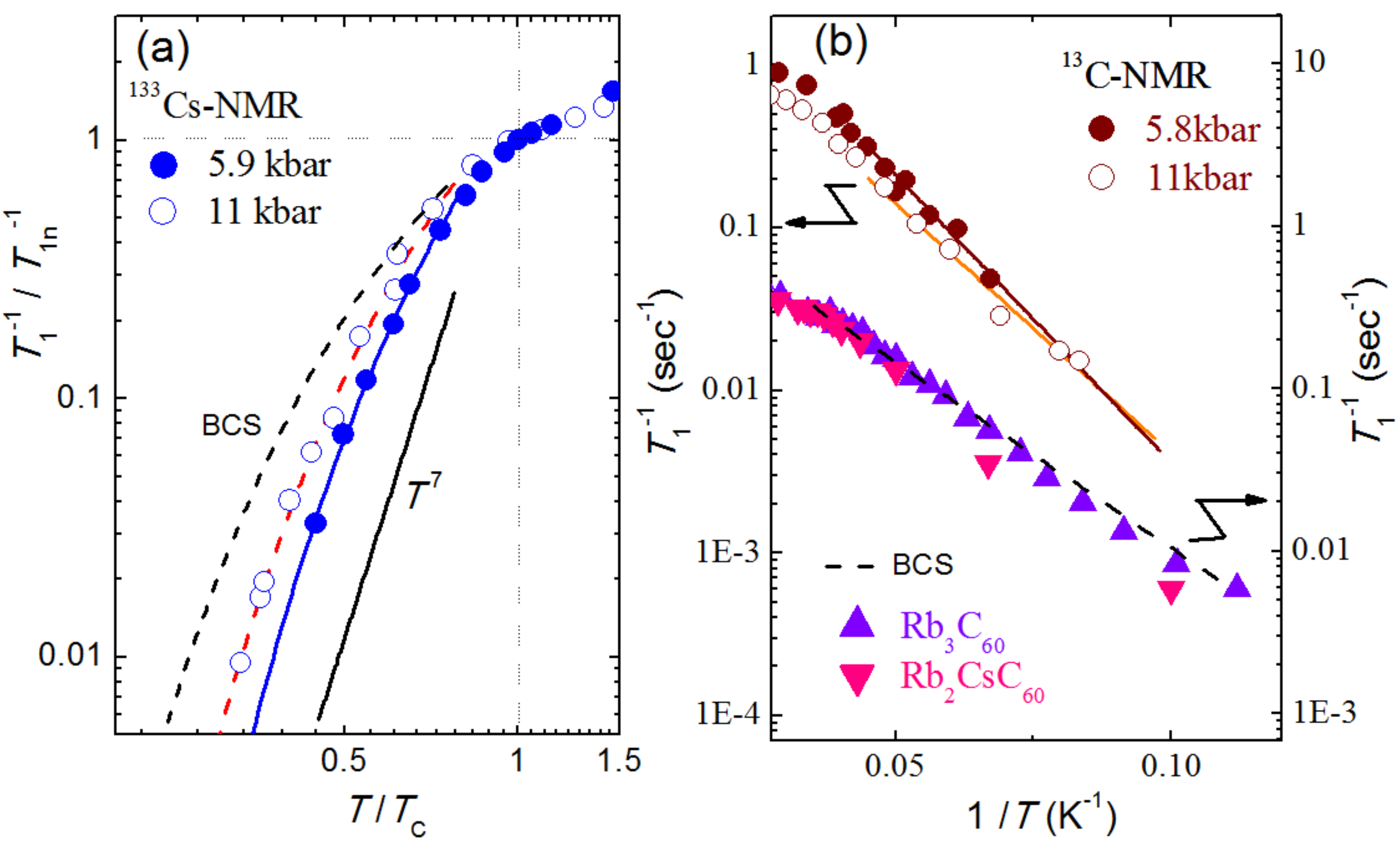} 
\end{center}
\caption{ (a) The behaviour of  $^{133}$ $T_{1}{}^{-1}$ in A15-Cs$_{3}$C$_{60}$ below $T_{c}$ would correspond to a power law exponent $T^{n}$ with  
$n\geq 7$(solid line) or to the exponential fits with 
$\Delta /k_{B}T_{c}=2.95$ (5.8 kbar) and $2.43$ (11 kbar). The index $n$ in $T_{1n}$ stands for the value taken at $T_c$. 
(b) The corresponding $^{13}T_{1}$ data (plotted in a logarithmic scale) also fit
this activated behaviour, with similar gap values ($\Delta /k_{B}T_{c}=2.7$ at 5.8 kbar and $2.2$ at 11 kbar). Data obtained in Rb$_{3}$C$_{60}$ \cite{Tycko-PRL68} and Rb$_{2}$CsC$_{60}$ \cite{pennington-RMP68} as well as the BCS slope
 ($\Delta /k_{B}T_{c}=1.75$) are shown for comparison. }
\label{Gap power law}
\end{figure}

$T_{1}$ measurements were conducted to better evaluate the SC gap. For both
 $^{13}$C and $^{133}$Cs, the recoveries of the nuclear magnetization are not
  exponential due to a weak distribution of relaxation
rates, similar to that found for the $^{13}$C NMR in
 K$_{3}$C$_{60}$ \cite{Holczer93epl}. This permits
us to determine an upper limit for $(T_{1})^{-1}$ \cite{sup}, which can be seen
to drop sharply in Fig.~\ref{Gap power law}(a),  of about a factor 30 at
 $\sim T_{c}/2$. The data in Fig.~\ref{Gap power law}(a) follow the
expected low-$T$ variation for a full $s$-wave gap $2\Delta $:
\begin{equation}
T_{1}^{-1}\propto (T_{1}T)_{n}^{-1}\exp (-\Delta /k_{B}T).
\label{gap rom T1}
\end{equation}
where $(T_{1}T)_{n}$ is the normal state value (approximately constant above $T_c$).
We therefore conclude that nodeless superconductivity common to the dense
fcc-\textit{A}$_{3}$C$_{60}$ systems still persists near $p_{c}$. Both fits
in Fig.~\ref{Gap power law}(a) correspond to $\Delta /k_{B}T_{c}\geq 2.5$,
quite larger than the BCS value of 1.75. Note that the lower 
$T$ data 
unambiguously correspond to \emph{longer} $T_{1}$ values than 
 expected for the  BCS gap: 
as discussed in \cite{sup} this cannot be explained by any experimental artifact or sample deficiencies.  

\begin{figure}[tb]
\begin{center}
\includegraphics[width=6.5cm]{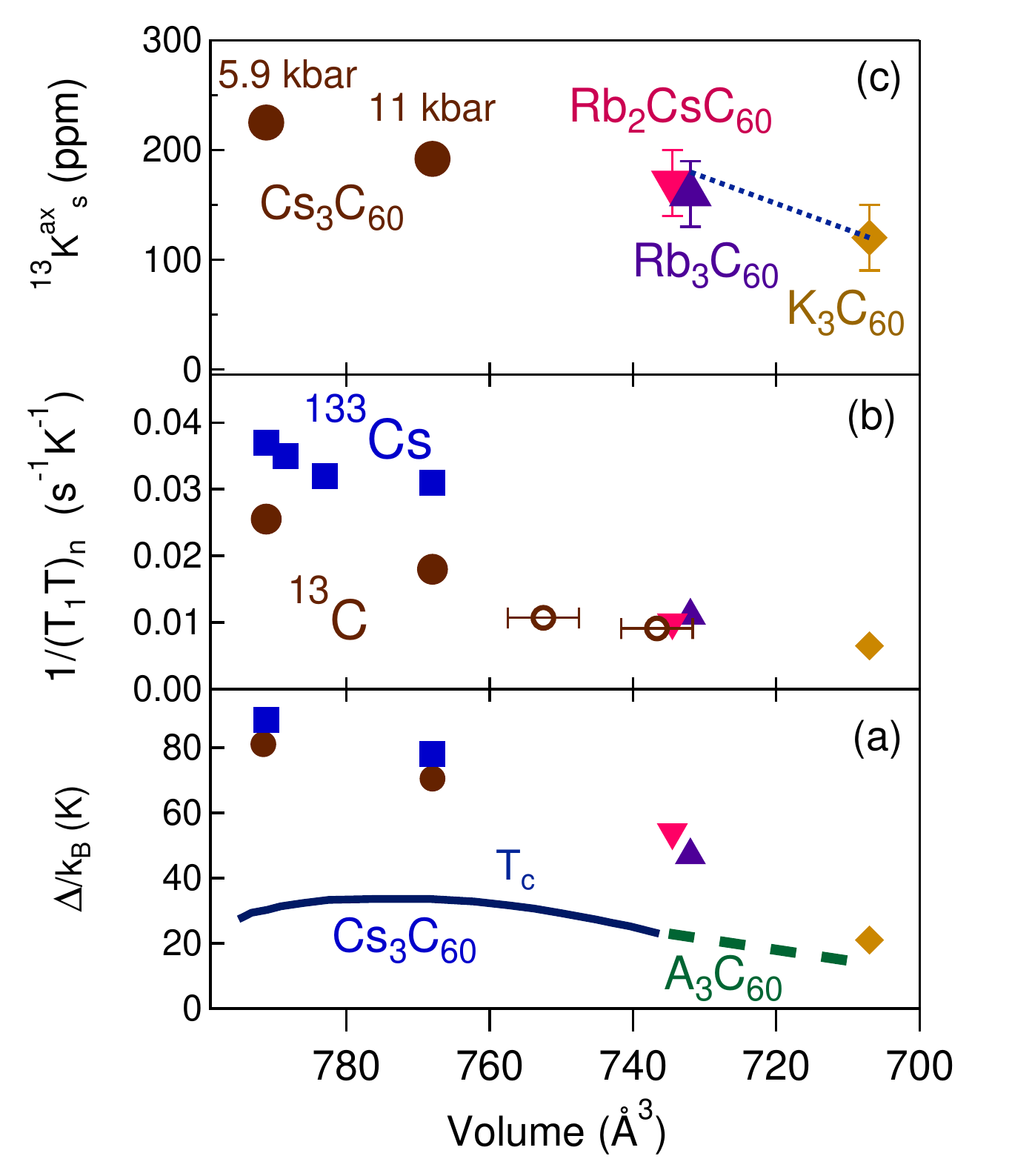} 
\end{center}
\caption{(a) The SC gap parameter $\Delta$ deduced from $T_{1}$ data of 
Fig.\protect\ref{Gap power law} increases regularly with $V_{C60}$,
 while $T_{c}$ goes through the SC dome (full line). The $T_{c}$ variation 
 for the dense fcc A$_{3}$C$_{60}$
compounds as determined from NMR \protect\cite{pennington-RMP68, Tycko-PRL68}
is shown in the dashed line. (b,c) Normal state  ($T=T_{c}$) values of (b) the 
$^{13}$C and $^{133}$Cs $(T_{1}T)^{-1}$ and of (c) the $^{13}$C anisotropic
Knight shift $K_{s}^{ax}$. In (b) the empty circles also show high $p$ $^{13}$C data from \protect\cite{Ihara-EPL94}.  The dotted line in (c) corresponds to the
variation of the ESR susceptibility (see text). The corresponding data for
other compounds are from \protect\cite{pennington-RMP68, Tycko-PRL68}. }
\label{Var with VC60}
\end{figure}

We illustrate in Fig.\ref{Gap power law}(b) that 
similar values for $\Delta$  are found for $^{13}$C NMR.
There we also show for comparison the $T_{1}$
data reported for the dense fcc alkali fullerides (Rb$_{2}$CsC$_{60}$
 \cite{pennington-RMP68} and Rb$_{3}$C$_{60}$ \cite{Tycko-PRL68}), for which $T_{c}$ are similar. 
 The results are summarized in Fig.\ref{Var with VC60}(a), where $\Delta $ is plotted 
 versus $V_{C_{60}}$, the volume per C$_{60}$ ball,
 so as to compare its variation with decreasing interball distance to that
obtained for $T_{c}$. There one can see that the gap increases continuously
with lattice expansion at variance with $T_{c}$ which goes through a
maximum. The ratio $\Delta /k_{B}T_{c}$ decreases progressively when $p>p_{c}$
 and approaches the BCS value of 1.75 only beyond the SC dome. The fast
drop of $(T_{1}{T})^{-1}$ below $T_{c}$  also points towards a
disappearance at $p_{c}$ of the Hebel-Slichter coherence peak \cite{sup}, 
seen in the dense $A_{3}$C$_{60}$ compounds \cite{pennington-RMP68,Ricco-PhC98,Kiefl-PRL70}.

\paragraph{Normal state spin susceptibility and $T_{1}$ data.}

$^{13}$C anisotropic shift data permit us to probe the variation
of the normal state $\chi _{s}$ versus $p$ upon approaching  $p_{c}$. 
We obtain  $^{13}K_{s}^{ax}$ by fitting 
 the spectrum above $T_c$ with $K_{orb}^{ax}=-120$ ppm \cite{note-Korb}. 
 We report in Fig.\ref{Var with VC60}(c) a small decrease of $^{13}K_{s}^{ax}(T_{c})$, 
that is of $\chi _{s}(T_{c})$, from $p=5.8$ to 11 kbar.
 There, the data for $^{13}K_{s}^{ax}$ in Rb$_{3}$C$_{60}$, Rb$_{2}$CsC$_{60}$
 and K$_{3}$C$_{60}$ 
 have been estimated from $^{13}$C spectra taken below 80K
 \cite{Tycko-PRL67,Yoshinari93prl,pennington-RMP68} for which rotation motions of
the C$_{60}$ balls are frozen and do not induce any line narrowing 
\cite{Yoshinari93prl}.
Such comparison is valid as the $^{13}$C hyperfine coupling 
is defined by the C$_{60}$ molecular properties independent of the compound. 
Finally, in Fig.\ref{Var with VC60}(c), the ESR
measurements of $\chi _{s}$ \cite{Robert98prb} (dotted line, arbitrary
units) give us as well an independent determination of the trend expected
for the dense $A_{3}$C$_{60}$.

The observed regular increase of 
$K_{s}^{ax}$ versus $V_{C_{60}}$ when approaching the Mott transition
contrasts with the well-established case of cuprates, for which the
occurrence of a pseudogap results in a large decrease of $\chi _{s}$ with
underdoping \cite{Alloul89prl}. Therefore a pseudogap cannot be
anticipated to occur in A15-Cs$_{3}$C$_{60}$.

We discuss now the pressure dependence of $R\equiv (T_{1}T)^{-1}$ in the
normal state just above $T_{c}$. The data versus $V_{C60}$ are summarized
in Fig.\ref{Var with VC60}(b) for $^{13}$C and $^{133}$Cs.  The variation of 
$^{13}R$  \cite{Ihara-EPL94}  parallels that for $\chi _{s}$, and corresponds
to a progressive increase from the smooth variation known  for the other 
fcc-$A_{3}$C$_{60}$ compounds \cite{maniwa-JPSJ63}. For $^{133}$Cs the relative
increase of $^{133}R$ for decreasing $p$ is somewhat smaller than for $^{13}$C.
Indeed $^{133}R/^{13}R$ increases regularly from $\sim 0.75$ in the Mott state at 1 bar
\cite{Jeglic-PRB80,Ihara-EPL94} to $\sim 1.4$ at 5.9 kbar and
 $\sim 2$ at 11~kbar.

The NMR relaxation rate $1/T_{1}$ probes the wave vector $\mathbf{q}$
dependent dynamic spin susceptibilities $\chi (\mathbf{q},\omega )$
according to the well-known Moriya relation 
$(T_{1}T)^{-1}$ $\propto \sum_{\mathbf{q}}A_{\mathrm{hf}}(\mathbf{q})^{2}\mathrm{Im}\chi (\mathbf{q},\omega)$.
 There the location of the probe nucleus with respect to the magnetic
sites determines the $\mathbf{q}$ dependence of $A_{\mathrm{hf}}(\mathbf{q})$.
The latter is $\mathbf{q}$ independent
for the $^{13}$C  spins, which probe the on-ball spin fluctuations, while
$^{133}A_{\mathrm{hf}}(\mathbf{q})$ is reduced at the AF wave vector for the 
$^{133}$Cs  spin, coupled to its 
neighbouring C$_{60}$. The progressive
decrease of $^{133}R/^{13}R$  with decreasing $p$ through the Mott
transition thus appears associated with a moderate increase of AF
fluctuations. The larger increase of $^{13}R$ with $V_{C_{60}}$ indicates
that both the AF fluctuations and the mass enhancement or increase of
density of states (DOS), responsible for the increase of $\chi _{s}$, do
contribute to the dynamic spin susceptibility.

\paragraph{Discussion.}

These experimental results permit us to address for the first time
experimentally the interplay between electron phonon coupling and electronic
correlations in a clean case. In a BCS formalism $T_{c}$ would be given by 
\begin{equation}
k_{B}T_{c}=1.14\hbar \omega _{D}\exp {\left( -1/\lambda \right) },
\label{eq:bcs}
\end{equation}
where $\lambda =V\rho (E_{F})$, $V$ being the electron-phonon coupling and 
$\rho (E_{F})$ the DOS at the Fermi level. The monotonic variation of
 $T_{c} $ versus $V_{C_{60}}$ (or lattice constant) found initially for the
dense fcc-$A_{3}$C$_{60}$ compounds has been used at length in the past to
indicate that the BCS formalism applies. This enforced the idea that the
Debye frequency $\omega _{D}$ and the electron-phonon coupling $V$ depend
solely on C$_{60}$ molecular properties, so that a smooth variation of
 $\rho (E_{F})$ with $V_{C_{60}}$ drives both variations of $T_{c}$ and
  $(^{13}T_{1}T)^{-1}$ \cite{gunnarsson-RMP69}. 
Here $T_{c}$ goes through a maximum versus $V_{C_{60}}$, while
 $(^{13}T_{1}T)^{-1}$ steadily increases,  which is indicative of a breakdown of 
 Eq.~(\ref{eq:bcs}).

The fast drop of $\chi _{s}(T)$ and $(T_{1}T)^{-1}$ for $T<T_{c}$ that we
found first reveals that the Hebel-Slichter coherence peak which is detected
for dense $A_{3}$C$_{60} $ is suppressed near the Mott transition \cite{sup}. 
Such a suppression is expected in the case of strong electron-phonon coupling.
But an increase near $p_{c}$ of this coupling cannot be expected, since it is a
molecular quantity independent of the proximity to the Mott state. This
departure from weak coupling BCS, together with the increase of
 $\Delta /k_{B}T_{c}$ with decreasing $p$ that we evidenced, should definitely be
associated with the growth of Coulomb correlations and the expected loss of
quasiparticle weight near $p_{c}$ \cite{Georges96rmp,Capone-RMP81}.

The Coulomb repulsion is usually partly taken into account in the BCS
formalism by assuming that the strength of the effective attractive coupling 
$\lambda $ gets decreased by a parameter $\mu ^{\ast }$ \cite{Scalapino66}.
An increase of $\mu ^{\ast } $ cannot explain a reduction of $T_{c}$ near
the Mott transition as it would necessarily drive the system back towards
weak coupling. On the other hand the increase of $\Delta /k_{B}T_{c}$  
is also found in strong-coupling BCS extensions where quasiparticle states
are renormalized by electron-phonon coupling 
\cite{Scalapino66}. Here, the observed increase of \emph{both} $\Delta $ and 
$\chi _{s}(T_{c})$ when $p$ decreases allows to think that the modification of
quasiparticle states affected by the increase of electronic correlations
near the Mott transition is important for the appearance of
superconductivity and the variation of $T_{c}$.

Such a possibility is apparently supported by the recent calculations by
Capone \emph{et al.} \cite{Capone-RMP81} who used DMFT to study the
interplay between Coulomb correlations and the attractive on-ball coupling.
As these authors point out, the effective coupling near the Mott transition
is governed by the interaction-renormalized bandwidth which vanishes at
 $p_{c}$ and therefore any BCS-like approach breaks down. This also results in
a large enhancement of  $T_{c}$ as compared to that solely expected from the
bare electron-phonon coupling and Eq.(\ref{eq:bcs}), so that electronic
correlations would help the SC state rather than suppress it. If such a scenario holds
 here, superconductivity might rather bear resemblance to local-singlet pairing schemes
developed for narrow-band superconductors \cite{Nozieres85lt}, where a $T_{c}$ dome is
also anticipated \cite{Micnas90rmp}. Note that the DMFT calculation of
Capone \textit{et al}. also allows for a dome-shaped behaviour of the gap
amplitude very close to the MI. The present experimental results should
definitely help to give some clue as to the relevance of such an approach if
refined computations of the gap, of $T_{c}$ and of the static and dynamic
susceptibilities can be performed.

\paragraph{Acknowledgements.}

We would like to acknowledge  Y. Ihara for his constant interest, and M.
Capone, M. Fabrizio and E. Tosatti for helpful discussions. We thank also
the ESF program INTELBIOMAT for supporting a short visit of T. Mito at LPS.

\bibliographystyle{apsrev4-1}
\bibliography{scCs3c60}

\begin{thebibliography}{32}
\bibitem [{not()}]{note-Korb}%
  \BibitemOpen
  \href@noop {} {}\bibinfo {note} {The anisotropic $^{13}$C NMR shape found in
  these compounds at low $T<80\ $K, for which the C$_{60}$ molecular rotations
  are frozen, correspond to $K_{orb}^{ax}$ values which vary from -110 to -120
  ppm with increasing charge between 0 and 4, so that $K_{orb}^{ax}\sim -120\
  $ppm can be taken for $n=3$.}\BibitemShut {Stop}%
\end{thebibliography}

\end{document}